\documentclass[conference]{IEEEtran}
\IEEEoverridecommandlockouts

\usepackage{cite}
\usepackage{amsmath,amssymb,amsfonts}
\usepackage{algorithmic}
\usepackage{graphicx}
\usepackage{textcomp}
\usepackage{xcolor}
\usepackage{booktabs}
\usepackage{multirow}
\usepackage{array}
\usepackage{url}
\usepackage{hyperref}
\usepackage{balance}
\usepackage{microtype}
\usepackage{subcaption}
\usepackage{tikz}
\usepackage{pgfplots}
\pgfplotsset{compat=1.18}
\usetikzlibrary{patterns}

\hypersetup{
    colorlinks = true,
    linkcolor  = blue,
    citecolor  = blue,
    urlcolor   = blue
}

\newcommand{\bvlm}{\textsc{BananaVLM}}
\newcommand{\binst}{\textsc{BananaInstruct}}
\newcommand{\pp}{\,pp}

\begin{document}

\title{BananaVLM: A Domain-Adapted Vision--Language Model for
Banana Crop Disease Diagnosis}

\author{%
\IEEEauthorblockN{Sangam Kumar Jena}
\IEEEauthorblockA{%
Department of Cyber-Physical Systems\\
Indian Institute of Science\\
Bangalore 560\,012, India\\
sangamjena@iisc.ac.in}
\and
\IEEEauthorblockN{Pandarasamy Arjunan}
\IEEEauthorblockA{%
Department of Cyber-Physical Systems\\
Indian Institute of Science\\
Bangalore 560\,012, India\\
samy@iisc.ac.in}
}

\maketitle

\begin{abstract}
Banana crop diseases are a major threat to food security across
Sub-Saharan Africa, South and Southeast Asia, and Latin America,
yet accurate field diagnosis remains challenging due to limited
expert availability and the visual similarity between disease
classes. Although recent vision--language models (VLMs) have
demonstrated strong general-purpose image understanding, they
systematically under-perform on fine-grained agricultural
recognition tasks. We present \bvlm{}, a domain-adapted VLM
for banana disease diagnosis built on LLaVA-v1.5-7B and
fine-tuned using Low-Rank Adaptation (LoRA). To address the
scarcity of multimodal agricultural instruction data, we
introduce \binst{}, a three-stage automated pipeline that
converts raw crop-disease image datasets into $\approx$80\,000
question--answer pairs without manual annotation. The pipeline
uses LLaVA-1.5-13B for image-grounded symptom description and
Mistral-7B for complex agricultural Q\&A synthesis and short-form
label grounding. Evaluated against 14 open-source and 5
closed-source VLM baselines on a nine-class banana disease
benchmark, \bvlm{} achieves 92.21\% in-domain and 83.28\%
out-of-domain (OOD) classification accuracy---surpassing the
best evaluated closed-source model (Gemini~2.5~Pro) by
$+$49.8\pp{} in-domain and $+$63.3\pp{} out-of-domain.
Binary healthy/diseased identification reaches 98.38\% OOD
with perfect recall (1.00). A controlled comparison with
Weight-Decomposed Low-Rank Adaptation (DoRA) reveals that its
magnitude--direction decomposition degrades multi-class
classification monotonically across epochs, confirming LoRA
as the superior adaptation strategy for fine-grained banana
disease tasks. Qualitative evaluation with nine diverse LLM
judges (G-Eval win rates 0.81--0.96) and blind pairwise
assessment by five domain experts (98.48\% win rate across
990 comparisons) further confirm that \bvlm{} generates
diagnostically accurate symptom descriptions and actionable
management recommendations far superior to any zero-shot
baseline. Our work establishes that lightweight, targeted domain
adaptation through automated instruction tuning is a scalable
and effective strategy for specialised agricultural AI.
Code, datasets, and model weights are publicly available at
\url{https://github.com/samy101/banana-vlm}.
\end{abstract}

\begin{IEEEkeywords}
vision--language models, crop disease diagnosis, banana, LoRA,
DoRA, instruction tuning, domain adaptation, agricultural AI,
multimodal learning
\end{IEEEkeywords}

%% ================================================================
\section{Introduction}
\label{sec:intro}
%% ================================================================

Agriculture underpins global food security, yet crop diseases
continue to cause yield losses of up to 40\% annually according
to the Food and Agriculture Organisation of the United Nations,
with the burden falling disproportionately on smallholder farmers
in low-income countries~\cite{mohanty2016using}. Banana is among
the most affected crops: a staple food and primary income source
for millions of smallholder farmers across Sub-Saharan Africa,
South and Southeast Asia, and Latin America, it is threatened by
a wide spectrum of diseases including Moko bacterial wilt, Black
Sigatoka, Panama wilt (Fusarium wilt), Cordana leaf spot,
Pestalotiopsis, Bract Mosaic Virus, and various insect pest
damages~\cite{bhuiyan2023bananalsd}. If not detected early,
diseases such as Moko and Black Sigatoka can devastate entire
plantations, causing irreversible damage to surrounding plants and
fields. Early and accurate disease diagnosis is therefore critical
not only for yield protection but also for directing appropriate,
cost-effective treatment before disease spread becomes irreversible.

Traditional diagnosis relies on expert agronomists---a resource
that is expensive, geographically constrained, and difficult to
scale in rural areas where extension service coverage is thin and
specialist consultations may require long travel times. In many
regions, farmers must wait days or weeks for a diagnosis, by which
point disease spread may have already caused severe damage.
Deep learning-based classifiers have shown strong
promise~\cite{mohanty2016using,too2019comparative}, but most
existing systems are label-only: they output a disease class
without explanations, symptom descriptions, or management
recommendations. Such opaque, ``black-box'' outputs are
insufficient for real deployment, since farmers and extension
workers require actionable guidance alongside any prediction, and
are unlikely to act on a prediction they cannot
interpret~\cite{shoaib2025revolutionizing}.

Vision--language models (VLMs) address this gap by combining
visual understanding with natural language generation. Models
such as LLaVA~\cite{liu2023llava} and its improved variant
LLaVA-1.5~\cite{liu2024improved} enable open-ended,
explanation-capable dialogue from a single image and
natural-language prompt---an interface that aligns naturally with
how agronomists communicate diagnostic findings, describing
visible symptoms, explaining diagnostic reasoning, and suggesting
contextually appropriate management strategies all in one
response. However, general-purpose VLMs trained on
internet-scale corpora consistently under-perform on fine-grained
agricultural tasks, as confirmed on benchmarks such as
AgMMU~\cite{gauba2025agmmu} and
AgroBench~\cite{shinoda2025agrobench}. The core problem is
distribution mismatch: subtle visual differences between disease
classes such as Sigatoka and Yellow \& Black Sigatoka, or
between early-stage Moko and healthy tissue, are rarely
represented in broad pretraining corpora and require
domain-specific fine-tuning to resolve.

Several recent efforts have explored domain-adapted VLMs for
agriculture. AgroGPT~\cite{awais2025agrogpt} demonstrated that
agricultural instruction corpora can be bootstrapped from
vision-only label-annotated data without requiring image--text
pairs. Agro-LLaVA-Next~\cite{xu2025agrollava} fine-tuned
LLaVA-NeXT on plant disease data, showing compelling results on
a single-crop benchmark. Despite this growing activity, no prior
VLM work has addressed banana crops comprehensively: existing
systems either address only a single or a few diseases, rely on
partially manual annotation, or lack rigorous out-of-domain (OOD)
evaluation against both open- and closed-source frontier
baselines.

We close these gaps with \bvlm{}, making the following
contributions:

\begin{enumerate}
  \item \textbf{\bvlm{}} --- a LoRA fine-tuned LLaVA-v1.5-7B
    model for nine-class banana disease diagnosis, achieving
    state-of-the-art performance on both in-domain and OOD
    benchmarks while remaining tractable for single-GPU training.
  \item \textbf{\binst{}} --- a fully automated three-stage
    pipeline producing $\approx$80\,000 multimodal Q\&A pairs
    from raw banana disease images without any manual annotation.
  \item \textbf{DoRA analysis} --- a controlled comparison of
    LoRA and DoRA under identical hyperparameters, revealing that
    DoRA's magnitude--direction decomposition degrades multi-class
    banana disease classification monotonically across epochs.
  \item \textbf{Comprehensive evaluation} --- systematic
    benchmarking against 19 baselines (14 open-source, 5
    closed-source), with dual-track qualitative assessment via
    G-Eval (nine LLM judges) and blind domain-expert pairwise
    evaluation (five agronomists, 990 comparisons).
  \item \textbf{State-of-the-art results} --- \bvlm{} surpasses
    Gemini~2.5~Pro by $+$49.8\pp{} in-domain and $+$63.3\pp{}
    OOD on classification, with human agronomists preferring
    \bvlm{} outputs in 98.48\% of blind comparisons.
\end{enumerate}

%% ================================================================
\section{Related Work}
\label{sec:related}
%% ================================================================

\subsection{Crop Disease Classification}

Early automated crop disease diagnosis relied on handcrafted
features---colour histograms, texture descriptors, and shape
features---combined with classical classifiers such as SVMs and
random forests~\cite{barbedo2013digital,phadikar2013rice}. The
PlantVillage benchmark~\cite{mohanty2016using} demonstrated high
CNN accuracy on controlled laboratory images, but performance
degraded substantially on field-captured
data~\cite{feueretal2018}, motivating domain adaptation, data
augmentation, and transfer learning from ImageNet-pretrained
models~\cite{too2019comparative}. Multi-crop datasets extended
evaluation beyond single-crop settings~\cite{thapa2020plant}.
Despite strong accuracy on benchmarks, CNN-based systems produce
only disease labels without explanations or management
guidance---a limitation documented in a comprehensive 2010--2024
review by Shoaib et al.~\cite{shoaib2025revolutionizing}, who
confirm that accuracy on benchmarks rarely translates to
interpretable or actionable outputs at the field level.

More recent multimodal work has begun to address this gap.
Liu et al.~\cite{liu2025crossmodal} proposed a cross-modal data
fusion architecture combining vision--language model descriptions
with image features via cross-attention, achieving better
performance than image-only models on soybean and PlantVillage
benchmarks. Wang et al.~\cite{wang2025llm} developed a
multimodal large language model for crop disease and pest
identification leveraging LoRA weight adaptation with a minimal
parameter increase. These works affirm the growing consensus that
label-only classification models are insufficient for real-world
agricultural deployment.

\subsection{Vision--Language Models}

Contrastive pretraining established the foundation for
visual--semantic alignment in models such as
CLIP~\cite{radford2021learning} and ALIGN~\cite{jia2021scaling}.
Generative VLMs---Flamingo~\cite{alayrac2022flamingo},
LLaVA~\cite{liu2023llava}, and
LLaVA-v1.5~\cite{liu2024improved}---extended this to
open-ended instruction-following. LLaVA-v1.5 introduced MLP
projectors and CLIP ViT-L/14-336 for improved visual resolution
and is adopted as the backbone of \bvlm{}. The open-source VLM
landscape has since expanded rapidly: Qwen-VL~\cite{bai2023qwenvl}
introduced strong grounding and multilingual capabilities, while
InternVL, MiniCPM-V~\cite{minicpmv2024}, and Gemma3~\cite{gemma2024}
progressively pushed multimodal reasoning performance. On the
closed-source side, the Gemini family demonstrates competitive
vision--language performance across diverse benchmarks. He et
al.~\cite{he2024specialist} showed that the performance gap
between general-purpose and task-specific VLMs represents a
fundamental limitation in specialised visual domains where
domain-specific knowledge is critical.

\subsection{Agricultural VLMs}

AgroGPT~\cite{awais2025agrogpt} bootstrapped agricultural
instruction data from vision-only annotated collections,
establishing the feasibility of expert-tuned agricultural VLMs
without reliance on image--text pairs.
Agro-LLaVA-Next~\cite{xu2025agrollava} fine-tuned LLaVA-NeXT on
plant disease images, demonstrating the effectiveness of
domain adaptation, but addressing only a single crop and lacking
rigorous OOD evaluation. Eleftheriadis et
al.~\cite{eleftheriadis2025selfconsistency} explored
self-consistency mechanisms for VLM-based precision agriculture
diagnosis, applying cosine-consistency self-voting to select the
most semantically coherent diagnosis from multiple candidates.
AgMMU~\cite{gauba2025agmmu} and
AgroBench~\cite{shinoda2025agrobench} provide agricultural
multimodal benchmarks that consistently expose the under-performance
of general-purpose VLMs on specialised agronomic tasks, directly
motivating the domain-adaptation approach of \bvlm{}.

\subsection{Parameter-Efficient Fine-Tuning}

Low-Rank Adaptation (LoRA)~\cite{hu2021lora} injects trainable
low-rank matrices into frozen attention layers, enabling
single-GPU adaptation of 7B-parameter models with dramatically
reduced memory requirements. Weight-Decomposed Low-Rank
Adaptation (DoRA)~\cite{liu2024dora} decomposes pretrained
weights into magnitude and direction components, updating
direction via a LoRA-style perturbation while keeping magnitude
as a separate trainable scalar, aiming to more closely replicate
the learning dynamics of full fine-tuning. Chen et
al.~\cite{chen2024dualora} proposed Dual-LoRA to separate skill
and task knowledge into distinct adapter modules, mitigating data
conflicts. We evaluate both LoRA and DoRA under identical
conditions in Section~\ref{sec:dora} and find that LoRA
substantially outperforms DoRA for this task.

\subsection{Automated Instruction Data Generation}

Prior VLM work used GPT-4 or GPT-4V to generate instruction data
from image captions and metadata~\cite{liu2023llava}.
Self-Instruct~\cite{wang2022self} established self-supervised
data generation as a general paradigm. A 2025 survey of
LLM-based synthetic data generation~\cite{maini2025synthetic}
identifies two primary strategies: caption-conditioned generation,
where image descriptions seed Q\&A synthesis, and
knowledge-injected generation, where external domain knowledge
enriches the supervision signal. Our three-stage \binst{} pipeline
combines both strategies. Gairola et al.~\cite{gairola2025multilingual}
demonstrated that instruction-tuned models trained on synthetic
agricultural Q\&A substantially outperform zero-shot baselines and
highlight the importance of human validation for domain-specific
evaluation---a methodology adopted in this paper.

%% ================================================================
\section{Dataset and \binst{} Pipeline}
\label{sec:data}
%% ================================================================

\subsection{Source Image Dataset}

The banana disease image dataset is derived from the Multi-Crop
Disease Dataset~\cite{multicrop2022}, a publicly available dataset
hosted on Mendeley Data. After removing corrupted or duplicate
images and standardising directory structure by disease label, the
dataset comprises 9,018 images across eight disease classes and one
healthy class (Table~\ref{tab:classwise}). The dataset exhibits
moderate class imbalance, with Yellow \& Black Sigatoka being the
most represented class (2,791 images) and Bract Mosaic Virus the
least (399 images), reflecting real-world prevalence distributions
in banana-growing regions. Class imbalance was preserved in
training so that the model learns realistic class priors.

\begin{table}[htbp]
\caption{Class-wise distribution of the \bvlm{} training set.}
\label{tab:classwise}
\centering
\setlength{\tabcolsep}{6pt}
\normalsize
\begin{tabular}{lc}
\toprule
\textbf{Disease / Condition} & \textbf{Images} \\
\midrule
Bract Mosaic Virus       & 399   \\
Cordana                  & 558   \\
Insect Pest              & 688   \\
Moko                     & 445   \\
Panama                   & 1,117 \\
Sigatoka                 & 1,368 \\
Yellow \& Black Sigatoka & 2,791 \\
Pestalotiopsis           & 573   \\
Healthy                  & 1,079 \\
\midrule
\textbf{Total}           & \textbf{9,018} \\
\bottomrule
\end{tabular}
\end{table}

\begin{figure}[htbp]
    \centering
    \includegraphics[width=\linewidth]{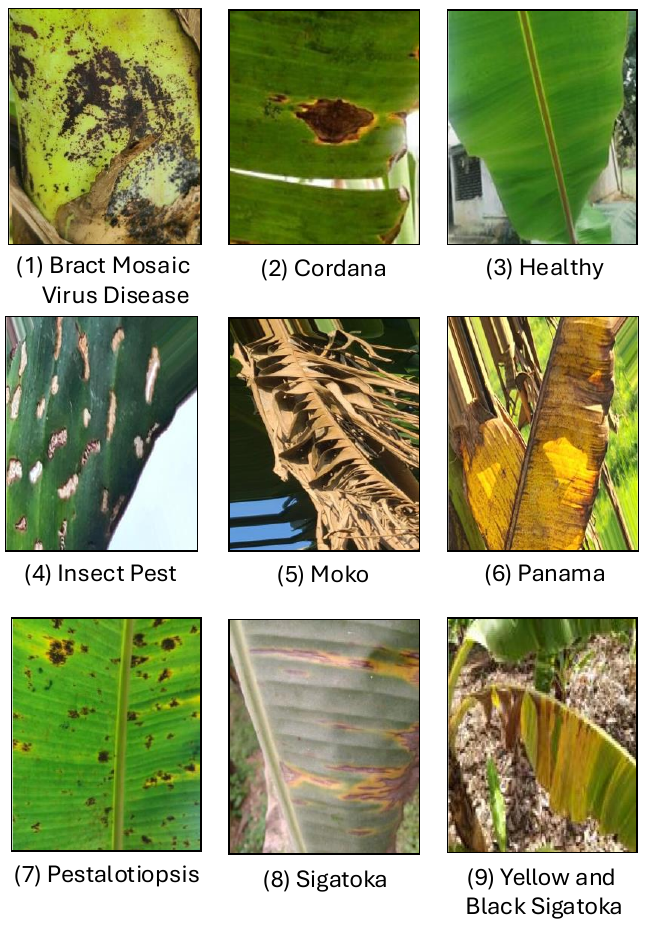}
    \caption{Representative images showing one sample from each
    of the nine banana disease/condition classes in the
    \bvlm{} training set.}
    \label{fig:dataset_overview}
\end{figure}

\textbf{Evaluation splits.} Two evaluation splits are used.
(i)~\emph{In-domain test set}: 905 images held out via stratified
10\% sampling from the same distribution as training, covering
all nine classes. Stratified sampling ensures proportional class
representation, preventing evaluation bias toward majority classes.
(ii)~\emph{Out-of-domain (OOD) test set}: 743 images from an
entirely separate banana disease dataset sourced independently
and not seen at any point during training or hyperparameter
selection, used to assess generalisation to images captured under
different conditions, with different cameras, and from different
geographic growing regions.

\subsection{\binst{}: Three-Stage Automated Pipeline}
\label{sec:pipeline}

\begin{figure}[htbp]
    \centering
    \includegraphics[width=\linewidth]{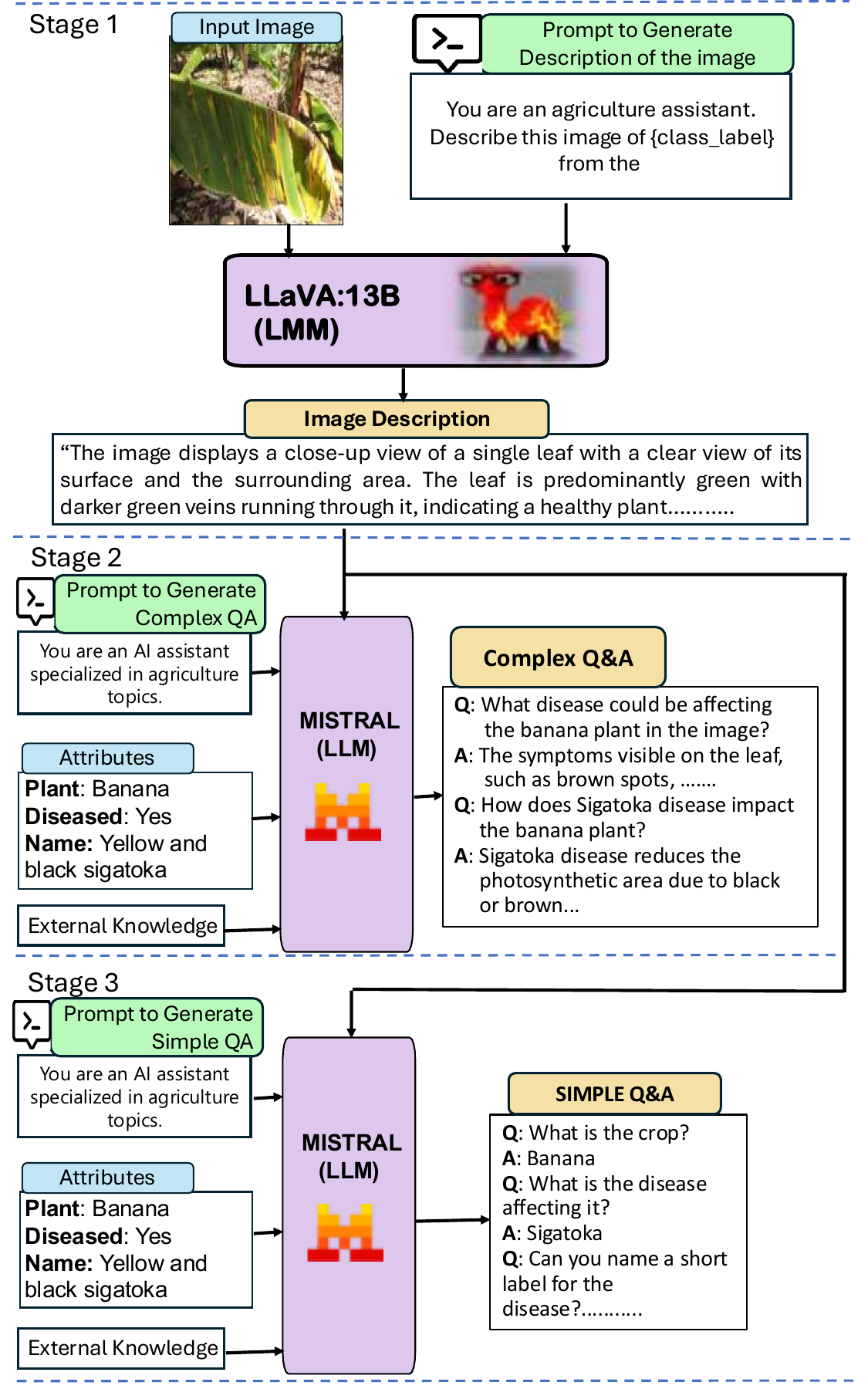}
    \caption{Three-stage \binst{} pipeline. }
    \label{fig:pipeline}
\end{figure}

\binst{} transforms raw image--label pairs into rich multimodal
instruction-tuning corpora without any manual annotation.
Rather than relying on costly expert annotation, the pipeline
uses a cascade of open-source language and vision--language models
to automatically generate diverse, agronomically grounded natural
language supervision from each image and its associated disease
label. Figure~\ref{fig:pipeline} illustrates the overall
framework, and Table~\ref{tab:prompts} shows the prompt templates
used at each stage.

The three-stage design is deliberate: Stage~1 anchors generated
language in observable visual evidence; Stage~2 builds domain
knowledge and reasoning ability on top of that visual grounding;
Stage~3 reinforces precise classification labels to reduce
inter-class confusion during fine-tuning.

\begin{table*}[htbp]
\caption{\binst{} pipeline prompt templates and outputs.}
\label{tab:prompts}
\centering
\setlength{\tabcolsep}{5pt}
\renewcommand{\arraystretch}{1.5}
\normalsize
\begin{tabular}{>{\bfseries}c l p{6.8cm} p{4.2cm}}
\toprule
\textbf{Stage} & \textbf{Model} & \textbf{Prompt Template} & \textbf{Output} \\
\midrule
1 &
  LLaVA-1.5-13B &
  \textit{``You are an agricultural assistant. Describe visible
  symptoms in this image of a \texttt{\{crop\}} plant affected by
  \texttt{\{disease\}}. Focus only on what is visible.''} &
  Image-grounded symptom description covering lesion morphology,
  chlorosis, necrosis, streak formation, and tissue collapse.
  Speculation and hallucination avoided by restricting to
  observable features only. \\
\midrule
2 &
  Mistral-7B &
  \textit{``[Stage~1 description] + [Disease attributes] +
  [External agricultural knowledge]. Generate 3--5 Q\&A pairs
  covering symptom interpretation, disease aetiology, progression
  dynamics, yield impact, climatic risk factors, and integrated
  management strategies. No speculation; no dataset references;
  no scientific names.''} &
  3--5 multi-turn Q\&A pairs per image; agronomist-style
  diagnostic and management dialogue spanning multiple agronomic
  dimensions. \\
\midrule
3 &
  Mistral-7B &
  \textit{``[Disease label + crop]. Generate 3--5 concise Q\&A
  pairs. Answers must be single words only. Focus on distinguishing
  visually similar disease pairs (e.g., Sigatoka vs.\ Yellow \&
  Black Sigatoka).''} &
  3--5 short-form Q\&A pairs for deterministic label grounding
  and fine-grained class discrimination; critical for substring
  matching evaluation. \\
\bottomrule
\end{tabular}
\end{table*}

\textbf{Stage~1 --- Image-Grounded Symptom Description.}
Each image is processed by LLaVA-1.5-13B, an open-source VLM
capable of fluent image-conditioned text generation without
internet access, making the pipeline fully self-contained and
reproducible. The crop name and disease label are injected into
the prompt as context, enabling agronomically precise vocabulary
while keeping outputs grounded in observable visual features.
Generated descriptions capture lesion morphology, chlorosis,
necrosis, streak formation, and tissue collapse while explicitly
avoiding speculation or unsupported claims.

\textbf{Stage~2 --- Complex Agricultural Q\&A.}
Mistral-7B~\cite{jiang2023mistral} receives the Stage~1
description alongside curated disease attributes (structured
per-disease metadata including affected plant parts, severity
indicators, and progression patterns) and external agricultural
knowledge (text passages from agricultural university extension
resources, national plant protection organisation documents, and
peer-reviewed articles). Between 3 and 5 Q\&A pairs are generated
per image, covering symptom interpretation, disease aetiology,
progression dynamics, yield impact, environmental and climatic
risk factors, and integrated disease management
recommendations---teaching the fine-tuned model to respond
meaningfully to a wide range of agronomist-style queries, not
merely to classify the disease label.

\textbf{Stage~3 --- Short-Form Label Grounding.}
Mistral-7B generates 3--5 concise Q\&A pairs to reinforce
deterministic label outputs, particularly important for
distinguishing visually similar pairs such as Sigatoka
vs.\ Yellow \& Black Sigatoka, where a model trained only on
conversational supervision might produce verbose or ambiguous
responses when a short classification label is expected. Answers
are constrained to single words or short phrases, ensuring the
model learns to produce clean, unambiguous outputs for direct
disease classification queries.

\textbf{Output and scale.}
After all three stages, the supervision for each image is
consolidated into a JSONL record containing: image file path,
Stage~1 symptom description, Stage~2 multi-turn Q\&A pairs, and
Stage~3 short-form label Q\&A pairs. With 3--5 Q\&A pairs per
image across three distinct supervision types, each image
contributes 7--11 distinct training interactions. The resulting
\binst{} corpus contains $\approx$80\,000 Q\&A pairs from 9,018
source images. A random subset of generated outputs for each
disease class was manually reviewed by agricultural domain experts
to verify factual consistency and agronomic correctness; only minor
corrections were required, confirming the pipeline does not
propagate systematic errors into training data. The complete
pipeline ran on a server equipped with an NVIDIA RTX~5090 GPU.

%% ================================================================
\section{Model Architecture and Training}
\label{sec:training}
%% ================================================================

\subsection{Architecture}

\bvlm{} is built on LLaVA-v1.5-7B~\cite{liu2024improved}, which
combines three components:

\begin{itemize}
  \item \textbf{CLIP ViT-L/14-336}: encodes the input image at
    336$\times$336 resolution into a sequence of patch-level
    visual tokens. The higher resolution (vs.\ standard 224) is
    important for agricultural disease diagnosis, where
    discriminative features such as lesion morphology, streak
    patterns, and chlorotic halos may span only a small region
    of the leaf surface. CLIP's contrastive pretraining on
    image--text pairs provides a useful initialisation for the
    agricultural domain even without domain-specific vision
    pretraining. This encoder is kept \emph{frozen} throughout
    all experiments.
  \item \textbf{Two-layer MLP projection head}: maps the visual
    token sequence from the vision encoder's embedding space into
    the language model's token embedding space. This projection
    head is \emph{not frozen}---it is fine-tuned alongside the
    LoRA weights to allow the model to adapt its visual-to-language
    mapping specifically to agricultural imagery and agronomic
    vocabulary.
  \item \textbf{Vicuna-7B language model}: based on
    LLaMA-2~\cite{touvron2023llama}, generates natural language
    responses conditioned on the concatenation of visual tokens
    (output by the projection layer) and the tokenised instruction
    prompt. Vicuna-7B's prior instruction tuning on conversational
    data gives it strong baseline capabilities for multi-turn
    dialogue, both of which are leveraged by the Stage~2 complex
    Q\&A supervision in \binst{}.
\end{itemize}

At inference time, the input image is passed through the CLIP
vision encoder to produce patch embeddings, which are projected
into the language model's embedding space by the MLP head and
prepended to the tokenised instruction sequence. Vicuna-7B then
attends jointly over visual and text tokens to generate the
target response autoregressively, enabling response to arbitrary
agronomic questions about the input image.

\subsection{LoRA Fine-Tuning}
\label{sec:lora}

Full fine-tuning of a 7B-parameter model requires prohibitive GPU
memory---typically in excess of 80\,GB of VRAM even with
mixed-precision training and gradient checkpointing. We therefore
adopt Low-Rank Adaptation~\cite{hu2021lora}, which introduces
trainable low-rank decomposition matrices into the frozen
attention layers of the language model while keeping all
pretrained weights fixed. The core insight is that task-specific
weight updates tend to have low intrinsic dimensionality:
\begin{equation}
  W = W_0 + \Delta W = W_0 + BA,
  \label{eq:lora}
\end{equation}
where $B \in \mathbb{R}^{d \times r}$, $A \in \mathbb{R}^{r \times k}$,
and $r \ll \min(d,k)$ is the LoRA rank. During training $W_0$ is
frozen; only $B$ and $A$ are updated. At initialisation $A$ is
drawn from a random Gaussian distribution and $B$ is set to zero,
so $\Delta W = BA = 0$ at the start of training and the model
begins from the pretrained checkpoint without disruption.

For $d = k = 4096$ and $r = 64$, the effective trainable
parameter count per LoRA layer is $r(d + k) = 524\text{K}$
versus $dk = 16.8\text{M}$ for full fine-tuning---a 32$\times$
compression per layer. Across all attention layers in Vicuna-7B,
the total trainable LoRA parameters represent a small fraction
of the full 7B model, enabling fine-tuning on a single 48\,GB
GPU. During inference the LoRA matrices can be merged back into
$W_0$ with no additional computational overhead.

LoRA weights are applied exclusively to the query, key, value,
and output projection matrices of the Vicuna-7B attention layers.
Gradient checkpointing is enabled throughout training to reduce
activation memory at the cost of a modest increase in compute.

\subsection{Training Configuration}

Table~\ref{tab:config} summarises the full training configuration.
Hyperparameters were determined through a combination of standard
practices for LoRA fine-tuning of 7B language models and
preliminary experiments on a held-out validation subset.

\begin{table}[htbp]
\caption{Training configuration for \bvlm{} (LoRA).}
\label{tab:config}
\centering
\setlength{\tabcolsep}{4pt}
\normalsize
\begin{tabular}{ll}
\toprule
\textbf{Parameter} & \textbf{Value} \\
\midrule
Base Model            & LLaVA-v1.5-7B \\
Vision Encoder        & CLIP ViT-L/14-336 (frozen) \\
LoRA Rank ($r$)       & 64 \\
LoRA Alpha ($\alpha$) & 16 \\
LoRA Dropout          & 0.05 \\
LoRA Target Modules   & q, k, v, o projections \\
Training Epochs       & 7 \\
Learning Rate         & $2 \times 10^{-5}$ \\
Projector LR          & $2 \times 10^{-5}$ \\
Batch Size / GPU      & 1 \\
Gradient Accumulation & 2 (eff.\ batch\,=\,2) \\
LR Scheduler          & Cosine \\
Warmup Ratio          & 0.03 \\
Weight Decay          & 0.01 \\
Max Sequence Length   & 1,024 tokens \\
Precision             & BF16 + TF32 \\
Hardware              & 1$\times$ NVIDIA RTX A6000 (48\,GB) \\
Optimiser             & DeepSpeed ZeRO-2~\cite{rajbhandari2020zero} \\
Checkpoint Strategy   & Per epoch (7 total) \\
\bottomrule
\end{tabular}
\end{table}

Several hyperparameter choices warrant justification. The rank
$r=64$ was chosen to balance representational capacity against
parameter efficiency: lower ranks (e.g.\ $r=8$ or $r=16$) are
common for general-purpose instruction tuning but may be
insufficient for fine-grained visual discrimination between disease
classes that differ in subtle ways. Setting $\alpha=16$ with
$r=64$ yields a scaling factor of $\alpha/r = 0.25$, which
moderates the magnitude of LoRA updates and prevents the low-rank
matrices from dominating the pretrained weight signal early in
training. The cosine decay schedule with warmup ratio 0.03
stabilises early training before the learning rate begins to
decay. Fewer than 2\% of training samples required truncation at
1,024 tokens, as the generated Q\&A pairs are designed to be
concise.

%% ================================================================
\section{Experiments and Results}
\label{sec:exp}
%% ================================================================

\subsection{Experimental Setup}

\textbf{Tasks.} Both tasks are evaluated using substring
matching between the model's generated text response and the
ground-truth disease label. A response is counted as correct
if the ground-truth label appears as a substring of the model's
output; this penalises responses that identify the wrong disease
or produce only generic descriptions while tolerating minor
variation in phrasing.

\begin{itemize}
  \item \emph{Identification} --- binary healthy vs.\ diseased
    detection. Recall is of particular importance, since a false
    negative (classifying a diseased plant as healthy) is
    considerably more costly in practice than a false positive.
  \item \emph{Classification} --- fine-grained nine-class
    disease prediction, providing the information required to
    select an appropriate treatment or management strategy.
\end{itemize}

\textbf{Baselines.} All 19 baselines are evaluated in a
zero-shot setting using the same standardised prompt that asks
the model to identify the disease and describe visible symptoms.
No few-shot examples are provided to any baseline.
Table~\ref{tab:testsets} summarises the per-class counts for
both evaluation splits.

\begin{table}[htbp]
\caption{Banana evaluation split statistics.}
\label{tab:testsets}
\centering
\setlength{\tabcolsep}{3pt}
\small
\begin{tabular}{lcc}
\toprule
\textbf{Class} & \textbf{In-Domain} & \textbf{OOD} \\
\midrule
Bract Mosaic Virus       & 40  & 50  \\
Cordana                  & 56  & 162 \\
Insect Pest              & 69  & 86  \\
Moko                     & 45  & 55  \\
Panama                   & 112 & 41  \\
Pestalotiopsis           & 58  & 173 \\
Sigatoka                 & 137 & --- \\
Yellow \& Black Sigatoka & 280 & 90  \\
Healthy                  & 108 & 86  \\
\midrule
\textbf{Total}           & \textbf{905} & \textbf{743} \\
\bottomrule
\end{tabular}
\end{table}

\subsection{Epoch-wise Training Dynamics}

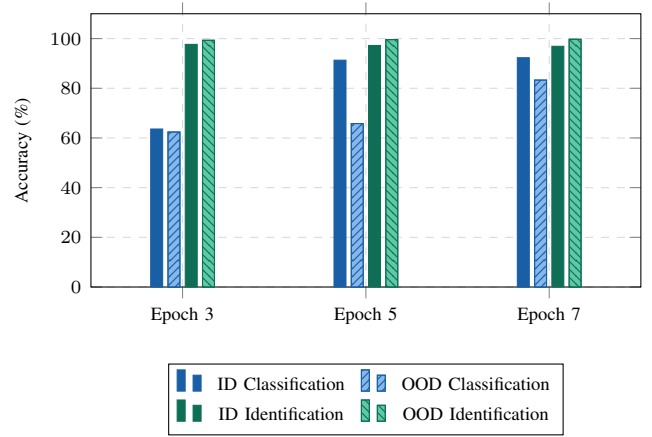
\begin{figure}[htbp]
\centering
\begin{tikzpicture}
\begin{axis}[
    ybar,
    bar width=4.5pt,
    width=\columnwidth,
    height=5.2cm,
    enlarge x limits=0.25,
    ylabel={Accuracy (\%)},
    symbolic x coords={Epoch 3, Epoch 5, Epoch 7},
    xtick=data,
    ymin=0, ymax=110,
    ytick={0,20,40,60,80,100},
    legend style={
        at={(0.5,-0.28)},
        anchor=north,
        legend columns=2,
        font=\scriptsize,
        column sep=4pt,
    },
    tick label style={font=\scriptsize},
    label style={font=\scriptsize},
    grid=major,
    grid style={dashed,gray!30},
    every node near coord/.append style={font=\tiny, rotate=90,
        anchor=west, inner sep=1pt},
]
%% ID Classification (dark blue)
\addplot[fill={rgb,255:red,24;green,95;blue,165},
         draw={rgb,255:red,24;green,95;blue,165}]
    coordinates {(Epoch 3,63.6)(Epoch 5,91.3)(Epoch 7,92.3)};
%% OOD Classification (light blue)
\addplot[fill={rgb,255:red,133;green,183;blue,235},
         draw={rgb,255:red,24;green,95;blue,165},
         postaction={pattern=north east lines, pattern color={rgb,255:red,24;green,95;blue,165}}]
    coordinates {(Epoch 3,62.4)(Epoch 5,65.8)(Epoch 7,83.3)};
%% ID Identification (dark teal)
\addplot[fill={rgb,255:red,15;green,110;blue,86},
         draw={rgb,255:red,15;green,110;blue,86}]
    coordinates {(Epoch 3,97.67)(Epoch 5,97.23)(Epoch 7,96.90)};
%% OOD Identification (light teal)
\addplot[fill={rgb,255:red,93;green,202;blue,165},
         draw={rgb,255:red,15;green,110;blue,86},
         postaction={pattern=north west lines, pattern color={rgb,255:red,15;green,110;blue,86}}]
    coordinates {(Epoch 3,99.32)(Epoch 5,99.59)(Epoch 7,99.73)};
\legend{ID Classification, OOD Classification,
        ID Identification, OOD Identification}
\end{axis}
\end{tikzpicture}
\caption{BananaVLM LoRA epoch-wise accuracy on in-domain (ID,
    solid fill) and out-of-domain (OOD, hatched fill) test sets.
  }
\label{fig:lora_epochs}
\end{figure}

LoRA classification accuracy improves monotonically across the
three evaluated checkpoints: 63.6\%\,$\to$\,91.3\%\,$\to$\,92.3\%
in-domain (epochs 3, 5, 7) and 62.4\%\,$\to$\,65.8\%\,$\to$\,83.3\%
OOD. The particularly large OOD gain between epochs~5 and~7
($+$17.5\pp{}) suggests the model continues consolidating
disease-specific visual representations beyond epoch~5, improving
generalisation to unseen imaging conditions. Identification
accuracy remains consistently high across all checkpoints
($>$96\% in-domain; $>$98\% OOD), indicating that the binary
healthy-vs.-diseased task converges early without degrading with
additional training. This asymmetry is expected: identification
requires a coarser visual signal (presence of disease-related
tissue damage), while classification requires fine-grained
discriminative features between classes sharing similar gross
visual appearance. All subsequent comparisons use the LoRA
checkpoint at epoch~7.

\subsection{Open-Source VLM Comparison}

Table~\ref{tab:opensource} reports identification and
classification accuracy for all 14 open-source baselines on both
splits.

\begin{table*}[htbp]
\caption{Open-source VLM comparison on in-domain and
  out-of-domain banana test sets.
}
\label{tab:opensource}
\centering
\setlength{\tabcolsep}{8pt}
\normalsize
\begin{tabular}{lcccc}
\toprule
\multirow{2}{*}{\textbf{Model}} &
  \multicolumn{2}{c}{\textbf{In-Domain}} &
  \multicolumn{2}{c}{\textbf{Out-of-Domain}} \\
\cmidrule(lr){2-3}\cmidrule(lr){4-5}
 & ID (\%) & Cls (\%) & ID (\%) & Cls (\%) \\
\midrule
LLaVA-7B~\cite{liu2023llava}                & 67.40 & 12.15 & 27.55 &  9.59 \\
LLaVA-13B~\cite{liu2023llava}               & 68.95 & 10.72 &  7.61 & 20.85 \\
LLaVA-34B~\cite{liu2023llava}               & 80.44 &  8.73 & 79.15 & 17.35 \\
Qwen2.5-VL-7B~\cite{wang2025qwen}           & 71.05 &  0.00 & 90.72 & 21.92 \\
Qwen3-VL-8B~\cite{qwen3vl2025}              & 85.97 & 12.04 & 99.24 & 21.46 \\
Qwen3-VL-32B~\cite{qwen3vl2025}             & 85.52 &  9.72 & 99.70 & 18.57 \\
Qwen2.5-VL-72B~\cite{wang2025qwen}          & 82.54 &  8.95 & 98.63 & 17.05 \\
LLaMA3-LLaVA-Next-8B~\cite{llavanext2024}   & 56.80 & 11.16 & 14.61 & 10.96 \\
Granite3.2-Vision~\cite{granitevision2024}  & 80.66 &  0.11 &  2.44 & 31.81 \\
Gemma3-4B~\cite{gemma2024}                  & 89.72 & 12.49 &100.00 & 14.92 \\
Gemma3-12B~\cite{gemma2024}                 & 90.94 &  4.86 & 99.70 & 15.22 \\
LLaVA-Phi3-3.8B~\cite{phi3technical2024}    & 17.57 &  2.32 & 11.87 & 27.70 \\
MiniCPM-V~\cite{minicpmv2024}               & 92.27 & 11.60 & 43.68 & 23.29 \\
BakLLaVA~\cite{bakllava2024}                & 21.66 &  0.44 & 17.20 &  0.00 \\
\midrule
\textbf{\bvlm{} (Ours)}  &\textbf{96.90}&\textbf{92.21}&\textbf{98.38}&\textbf{83.28}\\
\bottomrule
\end{tabular}
\end{table*}

Two patterns are evident across all baselines. First,
\emph{model scale does not predict classification accuracy}:
Qwen2.5-VL-72B, the largest open-source model evaluated, achieves
only 8.95\% classification in-domain---worse than several models
with less than one-tenth of its parameter count. Second,
\emph{high identification does not imply high classification}:
Gemma3-4B achieves 100\% OOD identification yet only 14.92\%
OOD classification, illustrating that detecting the presence of
visually abnormal tissue does not transfer to resolving \emph{which}
of nine disease classes is responsible. \bvlm{} exceeds the best
open-source classification baseline by $+$49.9\pp{} in-domain
and $+$51.5\pp{} OOD, confirming that domain specialisation
through targeted instruction tuning---not raw parameter
count---is the decisive factor for banana disease classification.

\subsection{Closed-Source VLM Comparison}

Table~\ref{tab:closedsource} compares \bvlm{} against five
Gemini model variants, all evaluated zero-shot via their
respective APIs.

\begin{table}[htbp]
\caption{Closed-source Gemini comparison on banana test sets.
  Best result per column in \textbf{bold}.}
\label{tab:closedsource}
\centering
\setlength{\tabcolsep}{4pt}
\normalsize
\begin{tabular}{lcccc}
\toprule
\multirow{2}{*}{\textbf{Model}} &
  \multicolumn{2}{c}{\textbf{In-Domain}} &
  \multicolumn{2}{c}{\textbf{Out-of-Domain}} \\
\cmidrule(lr){2-3}\cmidrule(lr){4-5}
 & ID (\%) & Cls (\%) & ID (\%) & Cls (\%) \\
\midrule
Gemini 2.5 Flash Lite   & 90.01 & 22.98 & 97.14 & 13.24 \\
Gemini 2.5 Flash        & 91.22 & 40.87 & 93.36 & 14.76 \\
Gemini 3 Flash Preview  & 93.90 & 40.08 & 96.44 & 23.90 \\
Gemini 3.1 Flash Lite   & 92.90 & 42.28 & 98.10 & 13.85 \\
Gemini 2.5 Pro          & 93.22 & 42.42 & 95.01 & 20.09 \\
\midrule
\textbf{\bvlm{} (Ours)} &\textbf{96.90}&\textbf{92.21}&\textbf{98.38}&\textbf{83.28}\\
\bottomrule
\end{tabular}
\end{table}

Gemini models achieve strong identification accuracy (90--99\%
in-domain; 93--98\% OOD), consistent with their general-purpose
visual understanding of vegetation and leaf tissue. However,
classification performance exposes a fundamental limitation of
zero-shot deployment: the best in-domain result among all Gemini
variants is 42.42\% (Gemini~2.5~Pro). Performance does not
improve monotonically with model tier---Gemini~3~Flash~Preview
reaches only 40.08\% despite being a newer generation---suggesting
classification accuracy on this task is not driven by general
capability improvements but by the absence of domain-specific
fine-grained knowledge that no amount of general pretraining
can supply. \bvlm{} surpasses Gemini~2.5~Pro by $+$49.8\pp{}
in-domain and $+$63.3\pp{} OOD on classification, while matching
or exceeding it on identification on every split. These results
demonstrate that LoRA fine-tuning on as few as $\sim$800 labelled
in-domain images injects banana-specific visual knowledge that no
evaluated zero-shot model of any scale approaches.

\subsection{Per-Class Analysis}

\begin{table}[htbp]
\caption{\bvlm{} per-class classification results (LoRA, epoch~7).
}
\label{tab:perclass}
\centering
\small
\setlength{\tabcolsep}{3pt}
\begin{tabular}{lccc|lccc}
\toprule
\multicolumn{4}{c|}{\textbf{In-Domain (797 imgs)}} &
\multicolumn{4}{c}{\textbf{Out-of-Domain (743 imgs)}} \\
\textbf{Class} & P & R & F1 & \textbf{Class} & P & R & F1 \\
\midrule
Bract Mosaic V.   & .97 & .95 & .96 & Cordana         & .99 & .92 & .96 \\
Insect Pest       & .90 & .44 & .59 & Pestalotiopsis  & .96 & .77 & .85 \\
Cordana           & .96 & .93 & .95 & Moko            & .95 &1.00 & .97 \\
Moko              &1.00 & .96 & .98 & Panama          & .74 & .90 & .81 \\
Panama            & .95 & .91 & .93 & Bract Mosaic V. & .67 & .98 & .80 \\
Pestalotiopsis    &1.00 & .88 & .94 & Yel.\&Blk.Sig.  & .57 &1.00 & .73 \\
Sigatoka          & .99 & .99 & .99 & Insect Pest     &1.00 & .25 & .40 \\
Yel.\&Blk.~Sig.  & .95 & .99 & .97 & & & & \\
\midrule
\multicolumn{4}{c|}{Overall: 92.21\%} &
\multicolumn{4}{c}{Overall: 83.28\%} \\
\bottomrule
\end{tabular}
\end{table}

Seven of eight classes achieve F1\,$\geq$\,0.93 in-domain, with
Sigatoka (F1\,=\,0.99) and Moko (F1\,=\,0.98) performing
strongest. The only underperforming class is Insect Pest
(F1\,=\,0.59), exhibiting low recall (0.44) despite adequate
precision (0.90): the model is conservative in predicting this
class---correct when it does predict it---but misses a substantial
fraction of true instances. This is consistent with Insect Pest
having limited training data (69 in-domain images) and high
visual overlap with healthy tissue, as insect damage in early
stages can manifest as minor discolouration without the
distinctive lesion patterns of fungal or bacterial diseases.

In-domain identification achieves 96.91\% accuracy with
\emph{perfect recall} (1.00): zero false negatives, meaning
no diseased plant is incorrectly cleared as healthy---the
highest-priority failure mode to avoid in real deployment.
OOD identification reaches 98.38\% with recall 1.00.

On the OOD set, Moko (F1\,=\,0.97) and Cordana (F1\,=\,0.96)
maintain near-in-domain performance on entirely unseen images,
confirming robust, transferable visual representations for these
classes. Insect Pest remains the hardest OOD class (F1\,=\,0.40,
recall\,=\,0.25), expected given its in-domain weakness and the
high variability of insect damage patterns across different sensors
and conditions. Yellow \& Black Sigatoka shows an interesting OOD
profile: precision drops to 0.57 while recall reaches 1.00,
consistent with a mild class-prior bias from this class being the
largest in training (2,791 images).

\subsection{DoRA Analysis}
\label{sec:dora}

We evaluated Weight-Decomposed Low-Rank Adaptation
(DoRA)~\cite{liu2024dora} under identical conditions ($r=64$,
$\alpha=16$, dropout\,=\,0.05, same hardware and learning rate)
to determine whether the additional complexity of DoRA's
magnitude--direction decomposition yields any benefit over
standard LoRA for banana disease classification. DoRA decomposes
$W_0$ into a magnitude component $m$ and unit-normalised
direction $V$:
\begin{equation}
  W_0 = m \cdot \frac{V}{\|V\|_c},
  \label{eq:dora_decomp}
\end{equation}
where $\|\cdot\|_c$ denotes the column-wise norm, and then
adapts both:
\begin{equation}
  W' = (m + \Delta m) \cdot \frac{V + BA}{\|V + BA\|_c}.
  \label{eq:dora_update}
\end{equation}
The motivation is that LoRA couples magnitude and direction changes
through the same low-rank update, potentially limiting expressiveness;
DoRA decouples them, aiming to more closely replicate the
magnitude--direction trade-offs of full fine-tuning.

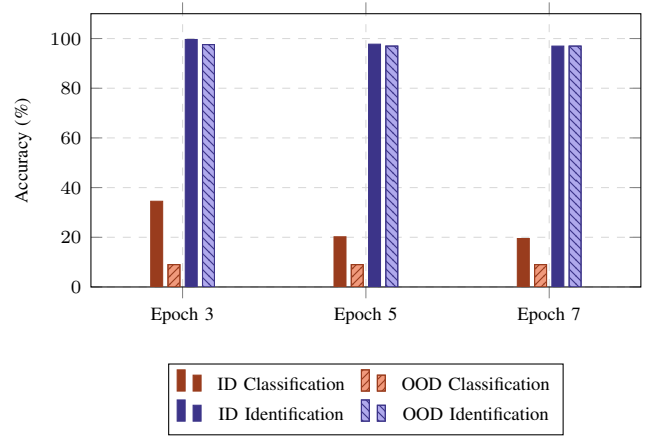
\begin{figure}[htbp]
\centering
\begin{tikzpicture}
\begin{axis}[
    ybar,
    bar width=4.5pt,
    width=\columnwidth,
    height=5.2cm,
    enlarge x limits=0.25,
    ylabel={Accuracy (\%)},
    symbolic x coords={Epoch 3, Epoch 5, Epoch 7},
    xtick=data,
    ymin=0, ymax=110,
    ytick={0,20,40,60,80,100},
    legend style={
        at={(0.5,-0.28)},
        anchor=north,
        legend columns=2,
        font=\scriptsize,
        column sep=4pt,
    },
    tick label style={font=\scriptsize},
    label style={font=\scriptsize},
    grid=major,
    grid style={dashed,gray!30},
]
%% ID Classification (dark coral/red -- degrading)
\addplot[fill={rgb,255:red,153;green,60;blue,29},
         draw={rgb,255:red,153;green,60;blue,29}]
    coordinates {(Epoch 3,34.5)(Epoch 5,20.2)(Epoch 7,19.5)};
%% OOD Classification (light coral -- near-flat ~9%)
\addplot[fill={rgb,255:red,240;green,153;blue,123},
         draw={rgb,255:red,153;green,60;blue,29},
         postaction={pattern=north east lines, pattern color={rgb,255:red,153;green,60;blue,29}}]
    coordinates {(Epoch 3,9.0)(Epoch 5,9.0)(Epoch 7,9.0)};
%% ID Identification (dark purple -- stable)
\addplot[fill={rgb,255:red,60;green,52;blue,137},
         draw={rgb,255:red,60;green,52;blue,137}]
    coordinates {(Epoch 3,99.62)(Epoch 5,97.73)(Epoch 7,96.96)};
%% OOD Identification (light purple -- stable)
\addplot[fill={rgb,255:red,175;green,169;blue,236},
         draw={rgb,255:red,60;green,52;blue,137},
         postaction={pattern=north west lines, pattern color={rgb,255:red,60;green,52;blue,137}}]
    coordinates {(Epoch 3,97.57)(Epoch 5,96.97)(Epoch 7,96.96)};
\legend{ID Classification, OOD Classification,
        ID Identification, OOD Identification}
\end{axis}
\end{tikzpicture}
\caption{BananaVLM DoRA epoch-wise accuracy on in-domain (ID,
    solid fill) and out-of-domain (OOD, hatched fill) test sets.
   }
\label{fig:dora_epochs}
\end{figure}

\textbf{Epoch-wise dynamics.}
Table~\ref{tab:dora_epochs} and Figure~\ref{fig:dora_epochs}
report DoRA results at epochs~3, 5, and 7. Three patterns are
evident. (1)~Identification remains intact ($>$96\% across all
epochs), confirming DoRA's decomposition does not catastrophically
disrupt the pretrained model's coarse visual understanding.
(2)~Classification \emph{degrades monotonically}: from 34.5\% at
epoch~3 down to 19.5\% at epoch~7 in-domain. (3)~OOD
classification hovers near 9\% throughout, unaffected by training
duration. This active degradation---rather than mere failure to
learn---distinguishes DoRA's failure mode from underfitting.

\begin{table}[htbp]
\caption{BananaVLM DoRA epoch-wise accuracy.
 }
\label{tab:dora_epochs}
\centering
\setlength{\tabcolsep}{4pt}
\normalsize
\begin{tabular}{ccccc}
\toprule
\multirow{2}{*}{\textbf{Epoch}} &
  \multicolumn{2}{c}{\textbf{In-Domain}} &
  \multicolumn{2}{c}{\textbf{Out-of-Domain}} \\
\cmidrule(lr){2-3}\cmidrule(lr){4-5}
 & ID (\%) & Cls (\%) & ID (\%) & Cls (\%) \\
\midrule
3 & 99.62 & 34.5 & 97.57 & $\approx$9 \\
5 & 97.73 & 20.2 & 96.97 & $\approx$9 \\
7 & 96.96 & 19.5 & 96.96 & $\approx$9 \\
\bottomrule
\end{tabular}
\end{table}

\textbf{Failure mode.}
This degradation is consistent with DoRA's magnitude--direction
decomposition introducing conflicting gradient signals for
fine-grained class-boundary learning. When the adaptation target
requires learning fine-grained, class-specific weight directions
(as is the case for distinguishing visually similar disease
classes), small errors in the separately trained magnitude
component $m$ amplify misclassifications by rescaling the
direction update $V+BA$ against the correct class boundary. LoRA's
joint update of magnitude and direction through the same low-rank
matrix product provides a more coherent gradient signal for
multi-class discrimination.

\textbf{LoRA vs.\ DoRA summary.}
Table~\ref{tab:lora_dora} directly compares both methods at
epoch~7 under identical hyperparameters. LoRA surpasses DoRA by
$+$72.7\pp{} in-domain and $+$74.3\pp{} OOD classification at
equal or better identification accuracy, conclusively establishing
LoRA as the superior adaptation strategy for banana disease
classification.

\begin{table}[htbp]
\caption{LoRA vs.\ DoRA at epoch~7.
  Identical hyperparameters: $r=64$, $\alpha=16$, dropout\,=\,0.05.
  ID\,=\,identification; Cls\,=\,classification.}
\label{tab:lora_dora}
\centering
\setlength{\tabcolsep}{4pt}
\normalsize
\begin{tabular}{lcccc}
\toprule
\multirow{2}{*}{\textbf{Method}} &
  \multicolumn{2}{c}{\textbf{In-Domain}} &
  \multicolumn{2}{c}{\textbf{Out-of-Domain}} \\
\cmidrule(lr){2-3}\cmidrule(lr){4-5}
 & ID (\%) & Cls (\%) & ID (\%) & Cls (\%) \\
\midrule
DoRA         & 96.96 & 19.5          & 96.96 & $\approx$9 \\
\textbf{LoRA}& \textbf{96.90} & \textbf{92.21}
             & \textbf{98.38} & \textbf{83.28} \\
\bottomrule
\end{tabular}
\end{table}

\textbf{System metrics (LoRA, epoch~7).}
Table~\ref{tab:sysmetrics} reports inference-time resource usage
for \bvlm{} at the final checkpoint, measured on a single
NVIDIA RTX~6000 GPU.
The Q\&A task (classification, symptom description, and management
strategy) incurs slightly higher latency than the binary
identification task owing to longer generated sequences, while both
tasks remain well within the 48\,GB memory budget of the RTX~6000.

\begin{table}[t]
\caption{Inference system metrics for \bvlm{} (LoRA, epoch~7) on a single NVIDIA RTX~6000 GPU.}
\label{tab:sysmetrics}
\centering
\begin{tabular}{p{2.2cm}ccc}
\toprule
\textbf{Task} & \textbf{Latency} & \textbf{Throughput} & \textbf{GPU Mem.} \\
 & \textbf{(s/img)} & \textbf{(img/s)} & \textbf{(MB)} \\
\midrule
Q\&A (Cls., Symp., Mgmt.) & 12.04 & 0.083 & 15,644.6 \\
Classification \& ID & 8.83 & 0.113 & 14,844.6 \\
\bottomrule
\end{tabular}
\end{table}
%% ================================================================
\section{Qualitative Evaluation}
\label{sec:qual}
%% ================================================================

Quantitative accuracy metrics capture label correctness but not
the quality, completeness, or agronomic utility of generated
responses. A model outputting the correct disease name in a single
word scores identically to one providing a detailed, clinically
accurate symptom description with actionable management guidance,
yet the latter is far more useful to a practising agronomist.
We therefore apply two complementary qualitative evaluation
protocols.

\subsection{LLM-as-a-Judge Evaluation (G-Eval)}

We adopt G-Eval~\cite{liu2023geval} with a panel of nine
open-source LLM judges. Each judge is provided with: (1) a
description of the input image, (2) expert-verified ground-truth
annotations (disease identity, visible symptoms, management
recommendations), and (3) the model's generated response. Judges
score responses on a 1--5 Likert scale across four agronomically
meaningful dimensions independently, without knowledge of which
model produced the response:

\begin{itemize}
  \item \textbf{Dis.ID} --- Disease Identification: correctness
    and specificity of disease presence detection.
  \item \textbf{Cls} --- Disease Classification: accurate and
    unambiguous naming of the correct disease class.
  \item \textbf{Symp.} --- Symptom Description: accuracy of
    visible symptom description with appropriate agronomic vocabulary.
  \item \textbf{Mgmt} --- Management Recommendation: accuracy,
    practicality, and contextual appropriateness of treatment guidance.
\end{itemize}

Win rate is the fraction of pairwise comparisons where \bvlm{}
outscores the base LLaVA model on the combined score across all
four dimensions. The panel comprises nine diverse open-source
LLMs spanning three model families, four parameter scales, and
multiple training paradigms to ensure aggregated win rates are
not an artefact of any single judge's tendencies or
biases~\cite{gu2025llmjudge}.

\begin{table}[htbp]
\caption{G-Eval scores --- \bvlm{} vs.\ base LLaVA (1--5 Likert).
  Format: Base\,/\,\bvlm{}. Win\,=\,fraction where \bvlm{} outscores base.}
\label{tab:geval}
\centering
\small
\setlength{\tabcolsep}{3pt}
\begin{tabular}{lccccc}
\toprule
\textbf{Judge} & \textbf{Dis.ID} & \textbf{Cls} & \textbf{Symp.} & \textbf{Mgmt} & \textbf{Win} \\
\midrule
Llama~3.1       & 1.47/4.28 & 1.87/4.45 & 2.86/4.63 & 2.87/4.61 & 0.92 \\
Mistral         & 1.16/4.72 & 1.13/4.46 & 2.60/4.23 & 2.79/3.97 & 0.83 \\
Mistral-Small   & 0.72/3.70 & 0.80/3.57 & 2.17/3.95 & 2.93/3.75 & 0.89 \\
Gemma3-12B      & 1.17/3.70 & 1.01/3.61 & 2.03/3.68 & 2.25/3.70 & 0.96 \\
Qwen2.5-14B     & 0.81/3.87 & 0.55/3.56 & 1.48/3.86 & 2.75/3.81 & 0.90 \\
Qwen2.5-32B     & 1.27/3.53 & 0.98/3.30 & 1.95/3.57 & 3.03/3.66 & 0.82 \\
Qwen2.5-7B      & 1.33/3.68 & 1.07/3.18 & 2.18/4.01 & 2.47/3.58 & 0.81 \\
Qwen2.5-72B     & 1.19/3.56 & 0.62/3.20 & 2.14/3.66 & 2.89/3.85 & 0.87 \\
Mixtral         & 1.90/4.37 & 1.45/4.27 & 2.43/4.63 & 2.33/4.37 & 0.96 \\
\bottomrule
\end{tabular}
\end{table}

\bvlm{} is preferred by every judge on every dimension, with win
rates ranging from 0.81 (Qwen2.5-7B) to 0.96 (Gemma3-12B and
Mixtral). The absolute score improvements are largest on the
disease identification and classification dimensions---base LLaVA
scores as low as 0.55--1.90, while \bvlm{} consistently scores
3.18--4.72---consistent with the quantitative accuracy gap between
the fine-tuned and base models. Symptom description and management
recommendation also show substantial improvement ($+$1--2 points
on average), confirming that domain fine-tuning improves not only
label accuracy but the full agronomic quality of the generated
response.

\subsection{Human Expert Evaluation}

Five domain-expert agronomists reviewed 198 pairwise comparisons
each (990 total) in a fully blind protocol: response pairs were
presented without model labels, and evaluators were not informed
which response came from which model. All ground-truth references
available to evaluators were cross-verified by domain experts
prior to use. Evaluators were asked to select the response they
considered more accurate, more complete in symptom description,
and more agronomically actionable in management guidance.

\begin{table}[htbp]
\caption{Human expert evaluation --- \bvlm{} vs.\ base LLaVA
  (198 pairwise comparisons per evaluator; 990 total).}
\label{tab:human}
\centering
\setlength{\tabcolsep}{5pt}
\normalsize
\begin{tabular}{lccc}
\toprule
\textbf{Evaluator} & \textbf{Base} & \textbf{\bvlm{}} & \textbf{Win Rate} \\
\midrule
Evaluator 1 & 1/198  & 197/198 & 99.49\% \\
Evaluator 2 & 0/198  & 198/198 & 100.00\% \\
Evaluator 3 & 7/198  & 191/198 & 96.46\% \\
Evaluator 4 & 4/198  & 194/198 & 97.98\% \\
Evaluator 5 & 3/198  & 195/198 & 98.48\% \\
\midrule
\textbf{Aggregate} & \textbf{15/990} & \textbf{975/990} & \textbf{98.48\%} \\
\bottomrule
\end{tabular}
\end{table}

Human experts preferred \bvlm{} in 975 of 990 comparisons
(98.48\%). One evaluator awarded a perfect win rate of 100\%.
The lowest individual win rate across all five evaluators is
96.46\%, indicating near-universal preference for the domain-adapted
model regardless of individual evaluator. With five evaluators
each making 198 binary judgements, the probability of such high
win rates arising by chance is negligible, and the consistency
across evaluators (effectively 1.0 in the direction of preferring
\bvlm{}) confirms that preference is driven by genuine quality
differences rather than individual evaluator bias.

Evaluators consistently cited three dimensions on which \bvlm{}
responses were preferred: (1)~\emph{accuracy of disease
identification}---the fine-tuned model consistently named the
correct disease, whereas the base model frequently produced
generic descriptions without committing to a diagnosis;
(2)~\emph{specificity of symptom description}---\bvlm{} described
observable features such as lesion morphology, discolouration
patterns, and affected plant parts with agronomically precise
language; and (3)~\emph{actionability of management
recommendations}---\bvlm{} provided contextually relevant
treatment suggestions (specific fungicide classes, cultural
practices, sanitation protocols) rather than the generic
``consult an expert'' responses from the base model.

\subsection{Qualitative Example}

Table~\ref{tab:moko_comparison} illustrates a representative case
where \bvlm{} correctly identifies Moko disease and its
characteristic symptoms, while both Gemini~3.1~Pro and GPT-5.3
produce incorrect diagnoses. This example highlights the practical
value of domain fine-tuning: general-purpose VLMs, lacking
crop-specific training signal, default to superficially similar
but agronomically incorrect diagnoses, whereas \bvlm{} correctly
identifies the bacterial wilt pathogen responsible for Moko.

\begin{table}[htbp]
\centering
\caption{Qualitative comparison on a Moko disease case.
  Correct identifications shown in {\color{green!60!black}green};
  incorrect in {\color{red}red}.}
\label{tab:moko_comparison}
\setlength{\tabcolsep}{4pt}
\small
\begin{tabular}{>{\bfseries}p{0.18\columnwidth} p{0.75\columnwidth}}
\toprule
& \includegraphics[width=0.22\textwidth]{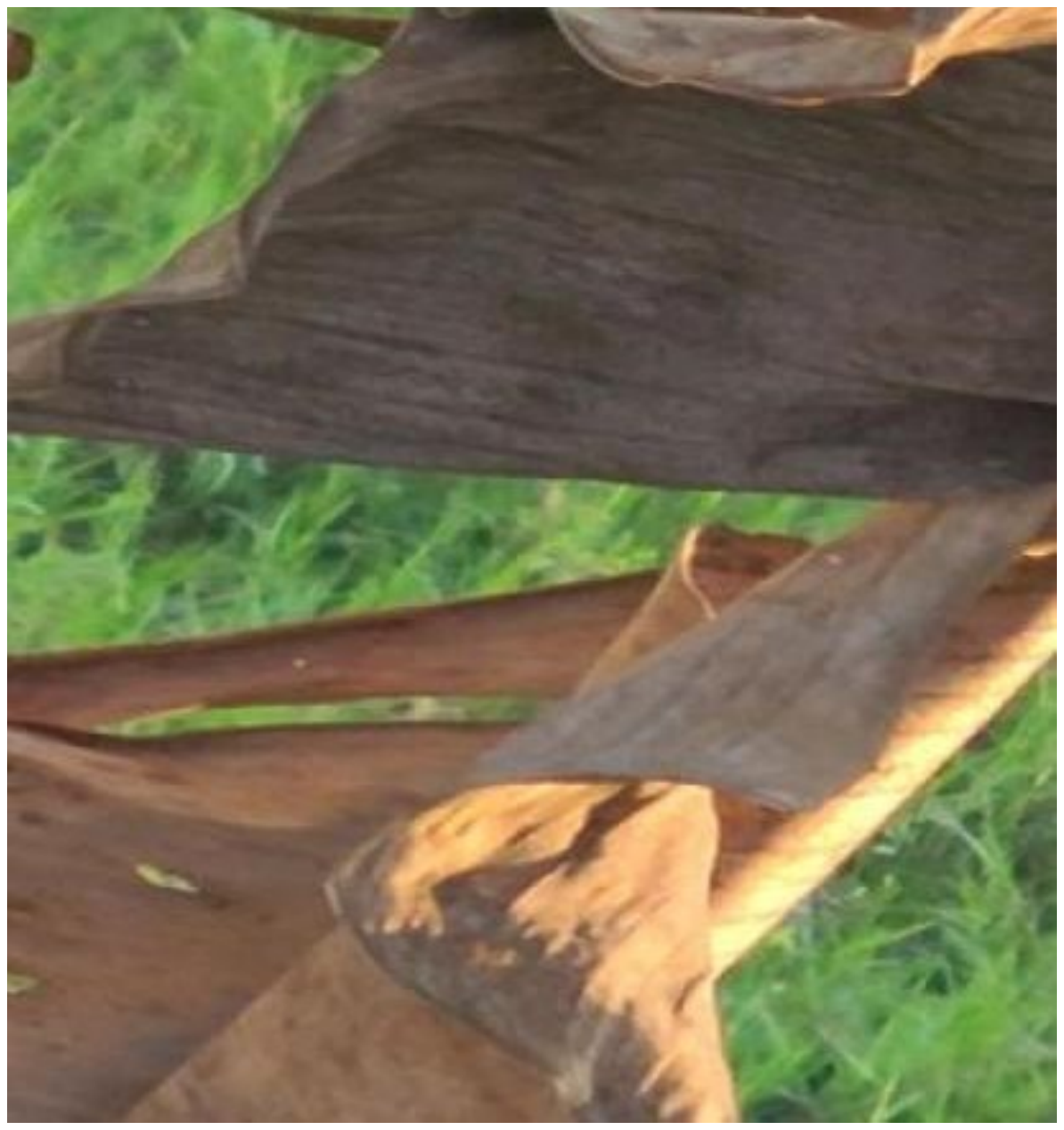} \\
\midrule
User &
Identify the disease and visible symptoms. \\
\midrule
Ground Truth &
\textbf{Disease}: MOKO.\newline
\textbf{Symptoms}: Leaf yellowing, wilting, vascular browning, fruit rot,
plant collapse.\newline
\textbf{Action}: Remove infected plants, disinfect tools, control insects,
use clean seedlings. \\
\midrule
\bvlm{} &
Classified as {\color{green!60!black}Moko}. Symptoms include
discoloration and deformation of leaves, which may turn
{\color{green!60!black}yellow or brown, become curled or twisted,
and exhibit wilting or drying out}. \\
\midrule
Gemini 3.1 Pro &
Most likely {\color{red}Sigatoka Leaf Spot (Yellow Sigatoka)}.\\
\midrule
GPT-5.3 &
Resembles {\color{red}severe leaf drying / senescence or
environmental damage} rather than a fungal or bacterial infection. \\
\bottomrule
\end{tabular}
\end{table}

%% ================================================================
\section{Discussion}
\label{sec:discussion}
%% ================================================================

\textbf{Domain adaptation decisively outperforms scale.}
The best zero-shot models---Qwen2.5-VL-72B (8.95\%),
Gemini~2.5~Pro (42.42\%)---achieve less than half of \bvlm{}'s
92.21\% in-domain classification accuracy despite having
far more parameters. Specialised instruction fine-tuning of
a compact 7B model consistently outperforms zero-shot prompting
of much larger systems, confirming that classification ability
is a function of domain-specific training signal rather than
general model capacity~\cite{he2024specialist}.

\textbf{Automated instruction generation scales VLM development.}
The entire $\sim$80K Q\&A corpus is generated without any manual
annotation, using only open-source models (LLaVA-1.5-13B and
Mistral-7B) and publicly available agricultural knowledge
resources. Model-collapse risk~\cite{maini2025synthetic}---a known
risk when models train iteratively on their own outputs---is
mitigated by using two different model families for generation
and by expert review of a random subset of outputs per disease
class. This pipeline is directly extensible to new crops,
lowering the barrier to agricultural VLM development.

\textbf{Identification--classification gap.}
Across all 19 zero-shot baselines, a consistent pattern emerges:
models can detect the coarser signal (presence of disease-related
tissue damage) but cannot resolve which disease class is
responsible. High identification accuracy does not imply high
classification accuracy (e.g., Gemma3-4B: 100\% vs.\ 14.92\%
OOD). This gap reflects the fundamental limitation of zero-shot
generalisation for fine-grained visual discrimination: without
seeing labelled examples of each disease class during training,
the model cannot reliably align disease-specific visual patterns
with their correct labels. \bvlm{} closes this gap entirely.

\textbf{LoRA vs.\ DoRA.}
DoRA's failure despite strong theoretical motivation is
practically important. Its magnitude--direction decomposition
introduces conflicting gradient signals for fine-grained
class-boundary learning in this multi-class setting, contrasting
with its reported benefits on NLP benchmarks~\cite{liu2024dora}.
This finding suggests that the DoRA decomposition is less
beneficial---or actively harmful---when the adaptation task
requires learning fine-grained visual class boundaries, and
practitioners should prefer LoRA for similar agricultural
fine-grained classification tasks.

\textbf{Limitations.}
Insect Pest remains the most challenging class, with F1\,=\,0.59
in-domain and 0.40 OOD, driven by limited training samples
(69 images) and visual overlap with healthy tissue. Yellow \&
Black Sigatoka exhibits a mild class-prior bias OOD from class
imbalance in training. These failure modes suggest avenues
for improvement through targeted data augmentation and
class-balanced sampling.

%% ================================================================
\section{Conclusion}
\label{sec:conclusion}
%% ================================================================

We presented \bvlm{}, a domain-adapted VLM for banana crop
disease diagnosis, and \binst{}, a fully automated three-stage
instruction generation pipeline. Fine-tuned on $\approx$80\,000
Q\&A pairs using LoRA on a single GPU, \bvlm{} achieves 92.21\%
in-domain and 83.28\% OOD classification across nine disease
classes---surpassing Gemini~2.5~Pro by $+$49.8\pp{} in-domain
and $+$63.3\pp{} OOD. In-domain and OOD identification both
achieve perfect recall (1.00), meaning zero false negatives at
the binary disease-detection level. Human agronomists preferred
\bvlm{} outputs in 98.48\% of 990 blind pairwise comparisons,
closely corroborated by G-Eval win rates of 0.81--0.96 across
nine independent LLM judges. A controlled DoRA comparison
confirms that LoRA is the superior adaptation strategy for
fine-grained banana disease classification, with DoRA exhibiting
monotonic classification degradation under identical
hyperparameters.

These results demonstrate that lightweight, targeted domain
adaptation through automated instruction tuning is an effective
and scalable strategy for specialised agricultural AI, enabling
a compact open-source 7B-parameter VLM to surpass proprietary
frontier models on specialised crop disease tasks.

\textbf{Future work.} Extension to wheat, tomato, and other
staple crops; multilingual voice-based deployment (a Hindi/English
prototype has already been developed); continual learning under
distribution shift; class-balanced sampling to address
under-represented disease classes; and large-scale field trials
with farmers and extension workers to evaluate real-world
robustness and usability.

Code, datasets, and model setup are at
\url{https://github.com/samy101/banana-vlm}.

\balance
\bibliographystyle{IEEEtran}
\bibliography{bananavlm_refs}

\end{document}